\documentclass[twocolumn]{aastex63}
\submitjournal{ApJ}

\usepackage{graphicx}% Include figure files
\usepackage{dcolumn}% Align table columns on decimal point
\usepackage{bm}% bold math
\usepackage{makecell}
\usepackage{verbatim} 
\usepackage{amsmath}

\begin{document}

%\preprint{Submitted to AJ}

\title{Comparative Planetology of Magnetic Effects in Ultrahot Jupiters: Trends in High Resolution Spectroscopy }

\correspondingauthor{Hayley Beltz}
 \email{hbeltz@umd.edu}

\author[0000-0002-6980-052X]{Hayley Beltz}

 \affiliation{Department of Astronomy, University of Maryland, College Park, MD 20742, USA}

 \author[0000-0003-3963-9672]{Emily Rauscher}
\affiliation{Department of Astronomy, University of Michigan, Ann Arbor, MI 48109, USA}

\begin{abstract}
Ultrahot Jupiters (UHJs), being the hottest class of exoplanets known,  provide a unique laboratory for testing atmospheric interactions with internal planetary magnetic fields at a large range of temperatures.  Thermal ionization of atmospheric species on the dayside of these planets results in charged particles becoming embedded in the planet's mostly neutral wind. The charges will resist flow across magnetic field lines as they are dragged around the planet and ultimately alter the circulation pattern of the atmosphere. We model this process to study this effect on high resolution emission and transmission spectra in order to identify observational signatures of the magnetic circulation regime that exist across multiple UHJs. Using a state-of-the-art kinematic MHD/active drag approach in a 3D atmospheric model, we simulate three different ultrahot Jupiters with and without magnetic effects. We post-process these models to generate high resolution emission and transmission spectra and explore trends in net Doppler shift as a function of phase. In emission spectra, we find that the net Doppler shift before and after secondary eclipse can be influenced by the presence of magnetic drag and wavelength choice. Trends in transmission spectra show our active drag models consistently produce a unique shape in their Doppler shift trends that differs from the models without active drag. This work is a critical theoretical step to understanding how magnetic fields shape the atmospheres of UHJs and provides some of the first predictions in high resolution spectroscopy for observing these effects. 
\end{abstract}

\section{Introduction}

Although astronomers have yet to directly detect a planetary magnetic field outside our solar system, there are many reasons to expect  exoplanets to host magnetic fields. In our own solar system, all of our gas giant planets host magnetic fields. Scaling laws based on geodynamo models \citep{Christensen2009} suggest the strength of internal dynamos should scale with the planet's bolometric flux, size of convective zone, and an efficiency factor that includes the temperature scale height. When applied to planets of similar size and orbit of hot Jupiters, this scaling law predicts dipole strengths ranging from 3 to 75 G \citep{Reiners2010}. Updating these scaling relationships with more accurate expectations of hot Jupiter internal luminosities give even higher values \citep{Yadav2017}.  Measurements of magnetic field strengths of brown dwarfs through radio emission indicate field strengths of several kG \citep{Kao2018}, but this technique has not yet been definitively accomplished for exoplanets \citep{Turner2021}.

Ultrahot Jupiters (UHJs) are an extreme class of exoplanets which experience strong irradiation and short orbital periods, often under three days. Due to these properties, these planets make excellent observational targets because of the favorable planet to star flux ratio they are able to provide. 
The physical environment of UHJs is  conducive for studying magnetic effects. 
The constantly irradiated dayside of the planet is hot enough to ionize atmospheric species  and many singly ionized metallic species have been observed (such as Ca II \citep{Casasayas2019, Yan2019, Borsa2021,Deibert2021, Taberno2021}, Fe II \citep{Hoeijmakers2018,  BenYami2020,Nugroho2020, Cabot2021, Kasper2021}, Mg II \citep{Sing2019, Stangret2022}, Si II \citep{Stangret2022}, Sr II \citep{Kesseli2021wasp76, Azvedosilva2022}, Ti II \citep{Hoeijmakers2018,Cauley2019,Prinoth2022,Jiang2023}, Sc II \citep{Merritt2021,Borsato2023}, Cr II \citep{Hoeijmakers2019,Belloarufe2023,Borsato2023}, V II \citep{Stangret2022}, Y II \citep{Hoeijmakers2019}, Ba II \citep{Azvedosilva2022,Jiang2023},and Tb II \citep{Borsato2023} for example)\footnote{see github.com/arjunsavel/hires-literature/ for an extensive list of HRS detections }.
%The constantly irradiated dayside of the planet is hot enough to ionize atmospheric species \textbf{(such as Ca II \citep{Casasayas2019, Yan2019, Nugroho2020kelt20, Borsa2021,Deibert2021, Taberno2021, Azvedosilva2022,Belloarufe2022,Kesseli2022w76spectralsurvey, Prinoth2022, ZhangY2022, Belloarufe2023,Borsato2023}, Fe II \citep{Hoeijmakers2018, Casasayas2019, Cauley2019, BenYami2020,Nugroho2020, Stangret2020, Borsa2021,Cabot2021, Kasper2021, Azvedosilva2022, Stangret2022, ZhangY2022, Belloarufe2023}, Mg II \citep{Sing2019, Stangret2022}, Si II \citep{Stangret2022}, Sr II \citep{Kesseli2021wasp76, Azvedosilva2022,Borsato2023}, Ti II \citep{Hoeijmakers2018,Cauley2019,Prinoth2022,Borsato2023,Jiang2023}, Sc II \citep{Hoeijmakers2019,Merritt2021,Borsato2023}, Cr II \citep{Hoeijmakers2019,Belloarufe2023,Borsato2023}, V II \citep{Stangret2022}, Y II \citep{Hoeijmakers2019}, Ba II \citep{Azvedosilva2022,Jiang2023},and Tb II \citep{Borsato2023} for example)\footnote{see github.com/arjunsavel/hires-literature/ for an extensive list of HRS detections }}. 
Combined with the strong winds predicted by many numerical models \citep{Showman2009,Rauscher2012GCM, DobbsDixon2013, KomacekShowman2016, Deitrick_2020THOR, Drummond2020, Helling2021, Steinrueck2021, Tan2024, Teinturier2024}, these moving ions will generate currents and time-changing electric fields, which would be influenced by a global planetary magnetic field \citep{Perna2010magdrag,Batygin2013}.  

Although these magnetic effects are expected to be present in exoplanets, only a small amount of numerical models have focused on these effects for UHJs. The most complex magnetic treatment are models that solve the full non-ideal MHD (magnetohydrodynamic) equations. In ``ideal" MHD, electric resistivity is considered to be negligible %\citep{Palenzula2009beyondidealmhd}. This results in 
and the Ohmic, Ambipolar and Hall diffusion terms are ignored  when solving the magnetic induction equation% \citep{Perna2010magdrag}.} 
However, the atmospheres of hot Jupiters should have non-negligible resitivities \citep{BatyginStevenson2010,Perna2010magdrag}, putting them within the ``non-ideal" regime. Non-ideal MHD models can show westward hotspot offsets or hotspot oscillation \citep{Hindle2021hotspotreversals}.  Additionally, It has been shown in MHD simulations on atmospheric columns of similar temperatures to the dayside of UHJ's that a  magnetic 'winding' effect caon occur, where the magnetic field strength will be increased until a steady-state is reached due to the high conductivity and strong zonal winds of the environment \citep{ClaudiaSoriano2023}. Full 3D non-ideal MHD models are scarce, but  \cite{Rogers2017} models the UHJ HAT-P-7b and provides a minimum field strength based on observed variability in the planet's hotspot offset over multiple \textit{Kepler} phase curves \citep{Armstrong2016}. 

Some atmospheric models apply a uniform drag timescale---that is, a single value for Rayleigh drag applied globally---to mimic the effects of magnetism \citep{KomacekShowman2016,Kreidberg2018,Arcangeli2019}. However, as pointed out in \cite{Beltz2022a}, this approach has non-physical implications when applied to UHJs. This is because applying a uniform drag to two locations that differ drastically in temperature, like the dayside equator and nightside equator, results in the unlikely scenario where the magnetic field is orders of magnitude stronger on the nightside than the dayside. A more physical method for approximating magnetic effects in 3D is our ''kinematic MHD" approach (described in more detail in section \ref{subsec: GCM}) which allows the drag timescale to vary globally and temporally, based on local temperature and pressure conditions. 

Recently, we published a set of three papers \citep{Beltz2022a,Beltz2022b,Beltz2023} exploring the effects of our kinematic MHD approach on the global circulation, high resolution emission spectra, and high resolution transmission spectra, respectively, on a model of the UHJ WASP-76b. In \citet{Beltz2023}, we proposed a set of observational predictions that could indicate the presence of a magnetic circulation regime. These predictions involved both high resolution spectroscopy (in trends of net Doppler shift vs phase) and phase curve behavior, the latter of which could potentially be observed by \textit{JWST}. In this work, we model two additional planets and WASP-76b again with an updated radiative transfer framework to determine how applicable the trends found in WASP-76b are to other UHJs. 

In Section \ref{sec: Methods}, we discuss the models presented in this paper and our kinematic MHD approach. We additionally discuss our post-processing technique and the method of obtaining net Doppler shifts from spectra. In Section \ref{sec: Results}, we explore the trends in net Doppler shifts across our 3 models in both high resolution emission and transmission spectra. We discuss the interpretation of these trends and important caveats in Section \ref{sec:Discussion}. Finally, we summarize our conclusions in Section \ref{sec:Conclusion}.

\section{Methods} \label{sec: Methods}
\subsection{GCM} \label{subsec: GCM}
We use a 3D numerical model which solves the primitive equations of meteorology (known as a GCM, General Circulation Model) in order to simulate the atmosphere of our planets. The GCMs in \cite{Beltz2022a,Beltz2022b,Beltz2023} all made use of a double-gray radiative transfer scheme. Under this scheme, radiation effects are calculated for two wavelength bands, known as the ``starlight" and ``thermal bands". A downside of this method are temperature profiles that are more isothermal than expected.  A more thorough explanation of the double-grey method can be found in \citep{Rauscher2012GCM,Lee2022RT}. For this paper, we model our planets using the more complex ``picket fence" method, which divides the radiative transfer into 5 bands, allowing for a more physical approach to radiative transfer. For details on how picket fence was implemented in our GCM and a comparison between double-gray and picket-fence models, see \citet{Lee2021, Malsky2023}.

In this work we  present models that have been treated with our ``kinematic MHD" or ``active magnetic drag" approach. This method, based on \citep{Perna2010magdrag}, was first applied to our GCM in \cite{RauscherMenou2013} for two hot Jupiters and was first applied to an UHJ in \cite{Beltz2022a}. More details about this scheme can be found in the mentioned papers, but in summary our drag timescale is a spatially and time varying Rayleigh drag. We assume that the planet's field is a dipole aligned with the axis of rotation, resulting in our drag being applied solely in the east-west direction. This timescale is calculated as follows: 
\begin{equation} \label{tdrag}
    \tau_{mag}(B,\rho,T, \phi) = \frac{4 \pi \rho \ \eta (\rho, T)}{B^{2} |sin(\phi) | }
\end{equation}
where $B$ is the chosen global magnetic field strength, $\phi$ is the latitude, $T$ is the local temperature and $\rho$ is the density for each model grid point. $\eta$ represents magnetic resistivity,
\begin{equation} \label{resistivity}
    \eta = 230 \sqrt{T} / x_{e} \textnormal{ cm$^{2}$ s$^{-1}$}.
\end{equation}
where ionization fraction, $x_{e}$, can be calculated using the Saha equation. For numerical stability, a minimum timescale of 0.0025 of the planet's rotational period  is imposed.
This method allows the model to vary the strength of the drag based on local conditions, which is more physically consistent than the uniform drag approach often used with GCMs. As discussed in \citep{Beltz2022a}, this approximation is valid for regions where the magnetic Reynolds number is $<1$, which is a vast majority of atmosphere modeled. The models approach this limit in the highest parts of the dayside upper atmosphere. Here, we can no longer safely assume that an induced magnetic field in this region is negligible compared to the surface field strength and a full non-ideal MHD approach is needed. Given that the vast majority of our atmosphere is not in this regime, we leave the addition of a full non-ideal approach for future work. 

\subsection{Planets Examined}
We use the GCM described above to model each UHJ of interest. These planets are some of the highest signal-to-noise observational targets and were chosen to survey a range of physical parameters present in the UHJ regime, specifically temperature, gravity, and rotation rate. Although these 3 planets cannot represent the entire UHJ parameter space, they offer informative sampling in this regime.  Each model was ran for at least 1000 days to allow the atmosphere to reach a quasi-equilibrium state. Sponge layers---model layers that include the addition of a Rayleigh drag to avoid the artificial reflection of waves off the model's  upper boundary---of varying strengths were used for each model, but were restricted to the top three layers of the atmosphere to not influence layers below the sponges. Each model assumed solar metallicity and an internal heat flux of 3500W.  The large scale planetary parameters are shown in Table \ref{tab:gcm_params} and the small differences between models regarding number of layers and pressure boundaries are described below.

\begin{deluxetable*}{lccccc}
\caption{UHJ Model Parameters} 
\label{tab:gcm_params}
\tablehead{ \colhead{Planet} & \colhead{Radius} & \colhead{Gravity} & \colhead{Orbital Period} & \colhead{Substellar Irradiation} & \colhead{Equilibrium Temperature}}
\startdata
         WASP-76b & $1.31 \times 10^{8}$ m  & 6.83 m s$^{-2}$ & 1.81 days & $5.14 \times 10^{6}$ W m$^{-2}$ & 2180 K\\ 
         WASP-121b &  $1.33 \times 10^{8}$ m & 8.45 m s$^{-2}$  &1.27 days & $7.01 \times 10^{6}$ W m$^{-2}$ & 2360 K \\
         WASP-18b &  $0.89 \times 10^{8}$ m & 166.28 m s$^{-2}$ & 0.94 days &  $8.02 \times 10^{6}$ W m$^{-2}$ & 2470 K
\enddata
\end{deluxetable*}

\subsubsection{WASP-76b}
For this work, we compare the double grey models of magnetic field strength 0~G and 3~G from \citep{Beltz2022a} with picket fence models of the same strength. This will be our only comparison to the double grey approach, so that we can determine if the trends found in \cite{Beltz2021,Beltz2022a,Beltz2022b} are robust under the picket fence modeling. All of these models extend from 100 bar to $10^{-5}$ bar over 65 layers, equally spaced in log $\sigma$ \footnote{$\sigma =$ local pressure / surface pressure, where the surface pressure  varies with latitude and longitude (as a result of mass conservation in the model), with an average value of 100 bars}. This planet has the lowest gravity and equilibrium temperature of the three presented. 

WASP-76b has been the target of many high resolution observational and theoretical studies. \cite{Kesseli2022w76spectralsurvey} provides an excellent exploration into the various species that have been detected in this planet's atmosphere. Extensive work has been done exploring the temperature structure for this planet \citep[such as][]{May2021,Gandhi2022,Beltz2022a,savel2022no}. One of the relevant datasets for this work are the net Doppler shifts found in \cite{Ehrenreich2020}. The data displays a ``kink'' or ``bottoming out'' behavior during the second half of transit, which was initially explained as a result of Fe cloud coverage. Multiple works with sophisticated models have tried to replicate this trend. \cite{Wardenier2021} use the MITgcm and decomposed the data to sectors of the planet to study the cross-correlation behavior of each limb and pole. They found the closest match by either artificially removing gaseous iron from the leading limb or by applying a larger temperature gradient between the limbs. \cite{savel2022no} additionally shows several sophisticated GCMs containing varying amounts of clouds and drag, but have difficulty reproducing the shape of this feature. By applying a non-zero eccentricity however they were able to match the magnitude of the Doppler shifts for part of transit. In \cite{Beltz2023}, we were able to reproduce the bottoming out structure by applying our active drag prescription to a cloud-free GCM of the planet. Although the magnitude of the trend was not reproduced, the shape of the trend matched remarkably well. 
\subsubsection{WASP-121b}
Of the three planets we present in this paper, WASP-121b is the ``in-between'' case. It has the second largest gravity, equilibrium temperature, and orbital period. Although not exactly the average of the two other planets, WASP-121b fills in our parameter space nicely. Here, we present picket fence models of 0G and 3G for the first time for this planet. Similar to the WASP-76b models, the models extend from 100 bar to $10^{-5}$ bar over 65 layers, equally spaced in log $\sigma$.

WASP-121b has been the subject of many high-resolution observations both in emission \citep{Hoeijmakers2022} and transmission \citep{BenYami2020,Bourrier2020,Cabot2020,Gibson2020,Hoeijamkers2020,Borsa2021,Merrit2021,Silva2022,Maguire2023} and has been the target of \textit{JWST} NIRSpec \citep{MikalEvans2023JWST}, \textit{Spitzer} \citep{Morello2023}, \textit{HST} \citep{MikalEvans2022HST} and TESS \citep{Bourrier2020} phase curves. Even in low resolution \textit{HST} phase curves, spectral differences between the day and night side are measurable \citep{MikalEvans2022HST}. Phase-resolved data shows hints of the transmission spectra becoming less blueshifted with time \citep{Ganhdi2023}. The variability of this planet has also been discussed \citep{Wilson2021, Changeat2024}
This planet has also been modeled with other GCMs \citep{Helling2021,Parmentier2018w121,Lee2022}, but never with any spatially varying drag.  %

\subsubsection{WASP-18b}
This planet has the highest gravity and equilibrium temperature of the three planets modeled. Due to the massive gravity of this planet, the models for this planet required a different setup for numerical stability purposes. These picket fence models extended from 100 bar to $10^{-3}$ bar over 50 layers. This choice of this upper boundary was due to numerical stability. We attempted to run models with boundaries of $10^{-5}$ bar and $10^{-4}$ bar, but the high gravity and temperature caused these models to  crash.   Given the high gravity of this planet, the IR photosphere is still included in our modeled pressure range, however it is possible that some line cores may be cut off. We discuss the potential implications of this upper boundary more in Section \ref{sec:Discussion}.   Surface field strengths of 0~G and 20~G were chosen as 20~G was the best fitting model when compared to \textit{JWST} phase curve and eclipse mapping observations \citep{Coulombe2023} The planet's high gravity makes it a poor high resolution transmission target, but emission observations have detected H$_{2}$O, CO, and OH \citep{Brogi2023W18,Yan2023}.

\subsection{A Note on Chosen Magnetic Field Strengths}
As described above, our treatment of magnetic drag makes some (reasonable) simplifying assumptions and requires a choice for the global/surface magnetic field strength. The drag caused by the chosen field strength is operating in tandem with other mechanisms affecting the flow structure such as hyperdissipation, metallicity, and sponge layers. And although the choice of field strengths for WASP-76b and WASP-18b are informed by \textit{Spitzer} and \textit{JWST} phase curves, we are not making a claim that these are the definitive values of the global field strength. Rather, in this work, \textit{we want to stress that we are exploring the observational differences between the magnetic and non-magnetic circulation regimes}, without claiming to know the specific field strength.

\subsection{Simulating Spectra}
We follow the radiative transfer routine outlined in \citet{Beltz2022b} and \citep{Beltz2023}. The methods for generating emission spectra are described in \citet{Zhang2017}. As in \citet{Beltz2022b}, emission spectra are calculated every 11.25$^{\circ}$, covering the entire orbit. 

The transmission routine is described in detail in \citet{Kempton2012} and \citet{savel2022no}. Following \citet{Beltz2023}, we calculate transmission spectra at 7 phases including mid-transit, separated by 3.75$^{\circ}$, or $\sim 0.01$ in phase.  

For each type of spectra, the GCM output containing temperature, east-west wind speed, and north-south wind speed at each grid location is interpolated onto an altitude grid in order to perform the necessary calculations. Spectra including broadening from winds and rotation (Doppler On spectra) as well as without (Doppler Off spectra) were calculated for each type and wavelength of spectra.  All of the spectra presented are calculated assuming solar abundances and local chemical equilibrium (LCE). We calculate spectra for two different wavelength ranges:
\begin{itemize}
    \item Wavelength 1: 2.3-2.35 $\mu$m; R=125,000;  Dominant Opacity source: CO
     \item Wavelength 2: 1.135-1.355 $\mu$m; R=125,000; Dominant Opacity source: H$_{2}$O

\end{itemize}
for direct comparison to \cite{Beltz2022b,Beltz2023}. Due to the high temperature of these planets, H$_{2}$O will dissociate on parts of the dayside, causing a non-uniform distribution of abundance around the planet. CO on the other hand will not dissociate even at these temperatures and thus will be more uniform in abundance. Recent work from \cite{Savel2023} suggests CO makes an excellent ''tracer molecule`` for this reason. The spectra generated at these two wavelength ranges thus will probe different atmospheric regions which will help inform how the combination of wavelength/opacity source and active magnetic drag influences spectra.

\subsection{Finding Net Doppler Shifts}
An extremely powerful aspect of HRS is the ability to detect the influence of winds \citep{Snellen2010}, and when data quality permits, determine the net Doppler shift of spectra as a function of phase. These net Doppler shifts give insight into the atmospheric flow, and we have previously shown \citep{Beltz2022b,Beltz2023} that these shifts could be used to infer the presence of magnetic fields. For HRS, the position of the line cores, which occur at shallower pressures than the continuum, will strongly influence the net Doppler shift. These line cores probe different physical regions of the planet compared to the continuum and our net Doppler shifts should be interpreted with this in mind.   The total net Doppler shifts will be influenced by a variety of processes including rotation, winds, wavelength choice, magnetic field strengths, and many more. These effects contribute in a multidimensional manner, further stressing the importance of 3D models. In order to calculate the net Doppler shift as a function of phase, we do the following: 
\begin{itemize}
    \item For each phase, calculate Doppler On and Off spectra, where the Doppler On spectra includes the broadening and shifting from winds and rotation. 
    \item Cross-correlate the two spectra. We use the same formalism as \cite{Brogi2019}. The cross-correlation coefficient at a shift of $x$ can be found:
    \begin{equation}
        C(x)=\frac{R(x)}{\sqrt{s_{On}^{2} s_{Off}^{2} }}
    \end{equation}
    where $s_{On}^{2}$ and $s_{Off}^{2}$ are the variance of the Doppler On and Off spectra respectively. $R(x)$ is the cross-covariance of the spectra:
    \begin{equation}
        R(x)= \frac{1}{N} \Sigma_{n} f(n) g(n-x)
    \end{equation}
    where $f$ is the Doppler Off  and $g$ is the Doppler On spectra. 
    \item Find the peak of the cross correlation curve and record the corresponding velocity shift. This is the net Doppler shift at this phase. 
\end{itemize}

Importantly, we calculate a pair of spectra for each phase, as opposed to using a single template Doppler Off spectra for the whole observation. Using a single template spectrum can bias the retrieved net Doppler shift by several km/s in emission spectra \citep{Beltz2022b}.

\section{Results} \label{sec: Results}
\subsection{Magnetic Circulation}
\begin{figure*}
    \centering
    \includegraphics[width=6in]{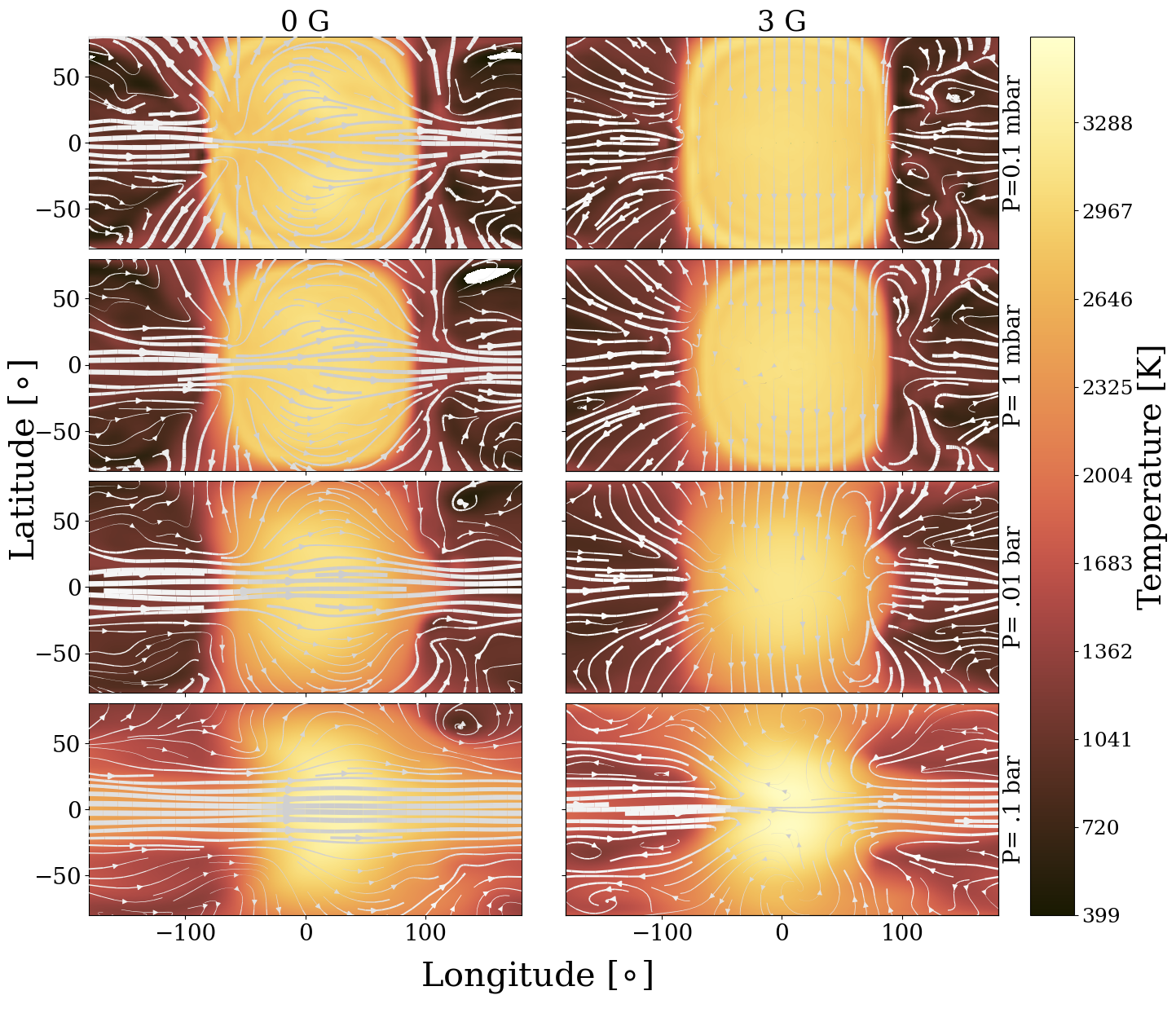}
    \caption{Maps of the temperature and wind structure at four pressure levels, in models of WASP-121b with (right) and without (left) active drag. The substellar point, corresponding to 0$^{\circ}$ longitude and latitude, is centered for each plot. The upper atmosphere of 3G active drag model of WASP-121b shows dayside flow moving poleward---in the north south direction---as opposed to the eastward motion seen in the drag-free model. We denote this pattern  as the ``magnetic circulation regime'' and note this behavior is seen across all planets modeled. }  
    \label{fig: W121stream}
\end{figure*}

Although the focus of this work is to understand trends in high-resolution spectroscopy, we will briefly touch on the GCMs themselves to get a broad picture of the atmospheric structure. For a much more in-depth exploration of the effect of our active magnetic drag on a UHJ, see \citet{Beltz2022a}.  This set of UHJ GCMs presented  show many similar features. In Figure \ref{fig: W121stream}, we show temperature and wind structure of the 0~G and 3~G model at a variety of pressures for WASP-121b as an example case. 

All models display strong day-night temperature gradients and the emergence of an equatorial jet on the nightside. For models including active drag, this jet is weakened at low pressures, resulting in small or no hotspot offsets, compared to the models without active drag which show significant offsets and a jet that persists through the dayside.  In addition, the application of our active drag brings about a magnetic circulation regime--first identified in \citet{Beltz2022a}--in the dayside upper atmosphere. This regime consists of dayside winds flowing poleward (in the north-south direction) as opposed to the mainly east-west flow seen in the drag free models. These different flow patterns alter the net Doppler shifts for both the emission and transmission spectra, as we will see shortly. 

\subsection{WASP-76b: Comparing Picket Fence and Double Gray}
\subsubsection{Atmospheric Structure}
\begin{figure*}
    \centering
    \includegraphics[width=5in]{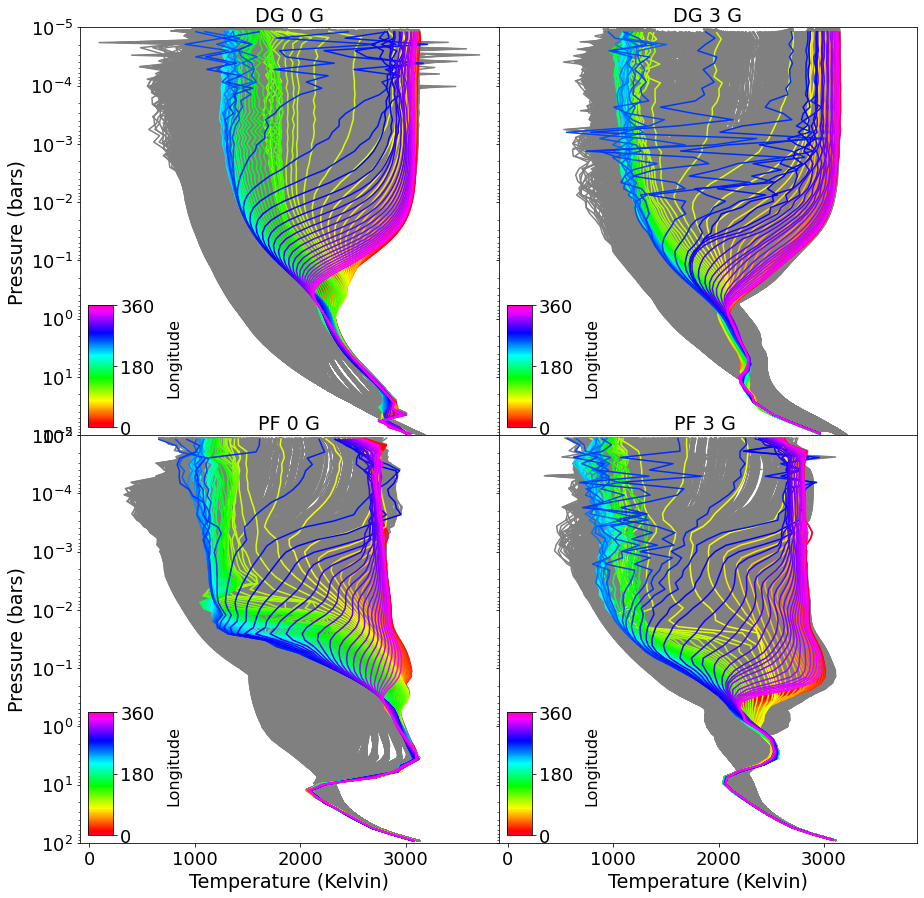}
    \caption{Temperature profiles of the four models of WASP-76b, with equatorial profiles labeled by color. The picket fence models are far less isothermal than the double-grey profiles. Additionally, the strength and location of thermal inversions are affected by the presence of our active drag, particularly for the picket-fence case.  } 
    \label{fig: W76DGPFTprof}
\end{figure*}

The atmospheric structure of the double gray models presented here is discussed at length in \cite{Beltz2022a}, so we will instead focus on the differences in structure between the double gray (DG) and picket fence (PF) models. While both models show the emergence of a magnetic circulation regime, important differences exist in the  wind and temperature structures which result in different predicted net Doppler shifts.  In Figure \ref{fig: W76DGPFTprof}, we show temperature-pressure profiles for the 4 cases of WASP-76b. When comparing the double-gray profiles to the picket-fence profiles, one can immediately see that the picket-fence profiles are less isothermal, regardless of the inclusion of active drag. The strengths and location of temperature inversions---both those caused by stratospheric absorbers and those dynamically produced--- also vary. In the double-gray approach, the strength of the temperature inversion caused by stellar irradiation is determined by the choice of two absorption coefficients (for these models, these coefficients chosen to best match the results from \citet{Fu2021}) while the picket-fence approach uses five coefficients. The use of more coefficients, which are calculated based on local conditions of the models, can thus create more physically realistic temperature inversions, which also vary in  location and extent. One might notice that the  PF substellar profiles appear to not have any temperature inversions in the upper atmosphere. We believe this is due to TiO/VO  dissociating at these temperatures, resulting in the upper atmosphere appearing isothermal.   

Comparing the two picket fence models, one can see that the 3~G model has a stronger dynamically produced deep (near $\sim 1$ bar) temperature inversion at the substellar point (corresponding to 0$^{\circ}/360^{\circ}$). Outside the substellar point, we see the dayside PF profiles showing inversions over a wide range of pressures with significant variation between the 3G and 0G case for both inversions due to stellar irradiation and dynamical interactions, the latter corresponding to the inversions near and below $\sim 1$ bar.  Both of the double gray models on the other hand have roughly the same inversion strengths and locations, due to the reduced number of coefficients in the DG approach. The picket fence coefficients on the other hand are more numerous and depend on local conditions, allowing for more coupling with the atmospheric thermal structure.  This highlights the flexibility of the picket fence model as it allows for more physically realistic inversions at a wider range of temperatures and pressures. 

For HRS, the presence---or lack thereof---of temperature inversions is a particularly important consideration, as the presence of a temperature inversion is directly detectable via the cross-correlation technique. Both the presence and the spatial extent of temperature inversions are influential, as different species will probe different atmospheric heights. The contrast between the temperature inversion structure in the DG and PF models suggests that when calculating high resolution spectra from a 3D model, the choice of radiative transfer will have a direct impact on the post-processed spectra and analysis. Thus, when possible, a more sophisticated radiative transfer routine should be used.
\subsubsection{Effects on Resulting Spectra}
\begin{figure}
    \centering
    \includegraphics[width=3.5in]{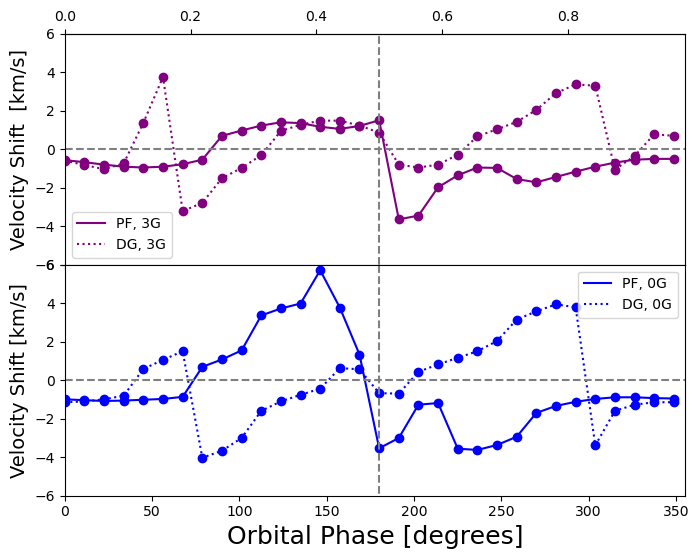}
    \caption{Comparison between double grey and picket fence net Doppler shifts near 2.3 micron. The dashed vertical line indicates the time of secondary eclipse. The models are most similar when the nightside of the planet is in view. Additionally, the radiative transfer routine choice is less impactful on the active drag models, which are more similar throughout orbit.  } 
    \label{fig: W76DGPF2_micron}
\end{figure}

As just shown above, the choice of radiative transfer routine will alter the temperature structure of the resulting model. This difference in temperature structure also results in  different 3D wind structures. We can see this influence in Figure \ref{fig: W76DGPF2_micron}, which compares the net Doppler shift of PF and DG models for emission spectra in the  2.3-2.35 $\mu$m range. The models are most similar when the nightside is in view. There are however significant differences between the predicted Doppler shifts in PF vs DG, particularly between first and third quadrature. These differences are the result of the complex 3D structure of the planet. 

\begin{figure*}
    \centering
    \includegraphics[width=6in]{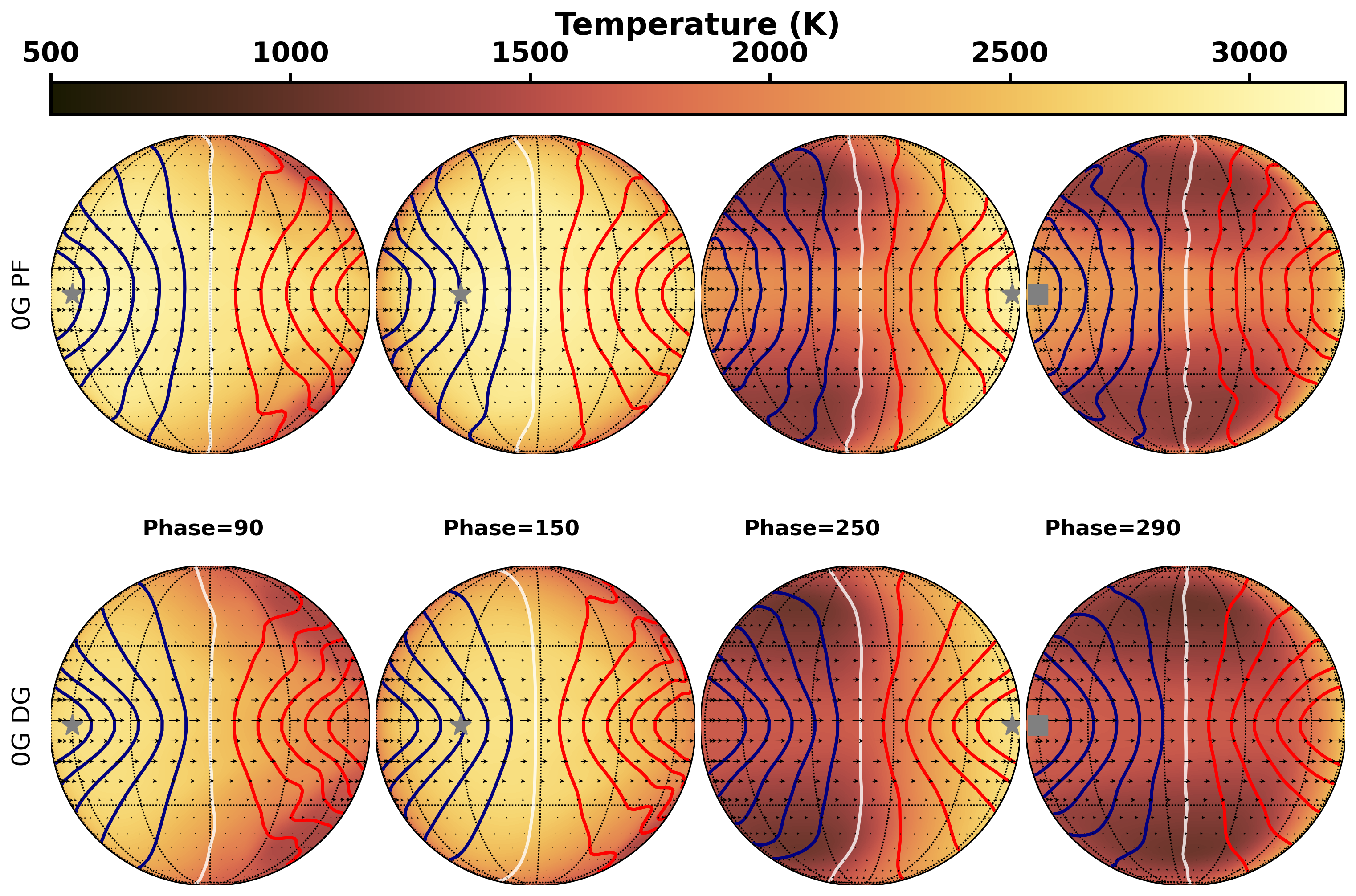}
    \caption{Orthographic projections of the 0G double gray and picket fence models near the IR photosphere at a variety of phases. The red and blue contours show lines of constant line of sight velocity in increments of 2 km/s. Between first quadrature and secondary eclipse, the PF model is more redshifted than the DG model and between secondary eclipse and third quadrature, the PF is more blueshifted for reasons discussed in the text.  }
    \label{fig:LOS W76DG+PF}
\end{figure*}

We include orthographic comparisons between the  0G models in Figure \ref{fig:LOS W76DG+PF}. We note that the shapes of the line of sight velocity contours in the IR photosphere are similar between both models, but the PF model has a higher temperature in the redshifted region of the planet between first quadrature and secondary eclipse (first two phases shown). This results in winds originating from hotter regions compared to the DG model, causing the PF model to be more redshifted for these phases. Around third quadrature (the last two phases), the planet has rotated such that part of the nightside is in view. Now, the blueshifted winds from the PF model overlap with hotter regions than the corresponding DG model, resulting in a stronger blueshift in the PF model.  Overall, the 3~G models differed less between radiative transfer scheme, likely due to the dominant wind structure being dictated by the active drag routine.  It should be noted that in HRS, the line strength is determined by the vertical temperature gradient. This introduces the possibility that a hot, vertically isothermal region may have less of an influence on the line position than a slightly cooler, less isothermal region. Thus, these photospheric maps should be combined with the temperature-pressure profiles shown in Figure \ref{fig: W76DGPFTprof} to gain a more complete picture on Doppler shifting behavior. Fundamentally, inspecting the temperature and wind patterns can give us some understanding of the Doppler signatures, but because this is an intrinsically complex, fully 3D radiative transfer problem, there are no simple ways to diagnose the outcome.  

\begin{figure}
    \centering
    \includegraphics[width=3.5in]{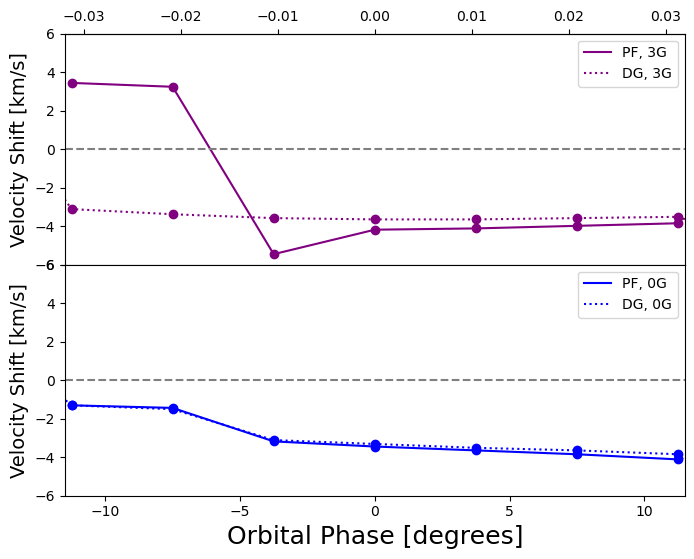}
    \caption{Comparison of net Doppler shifts between double grey and picket fence models for transmission spectra near 2.3 $\mu$m. These Doppler shifts show less variation between radiative transfer choice than the emission spectra.  } 
    \label{fig: W76DGPF2_micron_trans}
\end{figure}

We additionally explore how the choice of radiative transfer routine effects net Doppler shifts in high resolution transmission spectra. 
As shown in Figure \ref{fig: W76DGPF2_micron_trans}, the differences between the PF and DG models in transmission are smaller than in emission spectra, particularly for the 0G models.  This is due to the models having extremely similar line-of-sight velocity profiles at the east and west limbs during transit. (See  Appendix Figure \ref{fig:LOS W76DG+PFtransit}). The active drag 3G models differ more during the first half of transit, but are very similar during the second half.  However, in \cite{Beltz2023}, where we calculated spectra at a slightly larger phase range, we found that the 3G DG model is redshifted at a phase of $\sim -.04$ (shown in  Figure 6 of that work). Thus, both of the active drag models show the same sharp jump from being redshifted to blueshifted, just offset slightly in phase. 
Given that the other planets were modeled with the picket fence method, and that this method should produce more physically accurate temperature structures, we will restrict ourselves to corresponding WASP-76b models while we examine trends across multiple planets. 

\subsection{Trends across Planets: Emission Spectra}
\begin{figure*}
    \centering
    \includegraphics[width=6in]{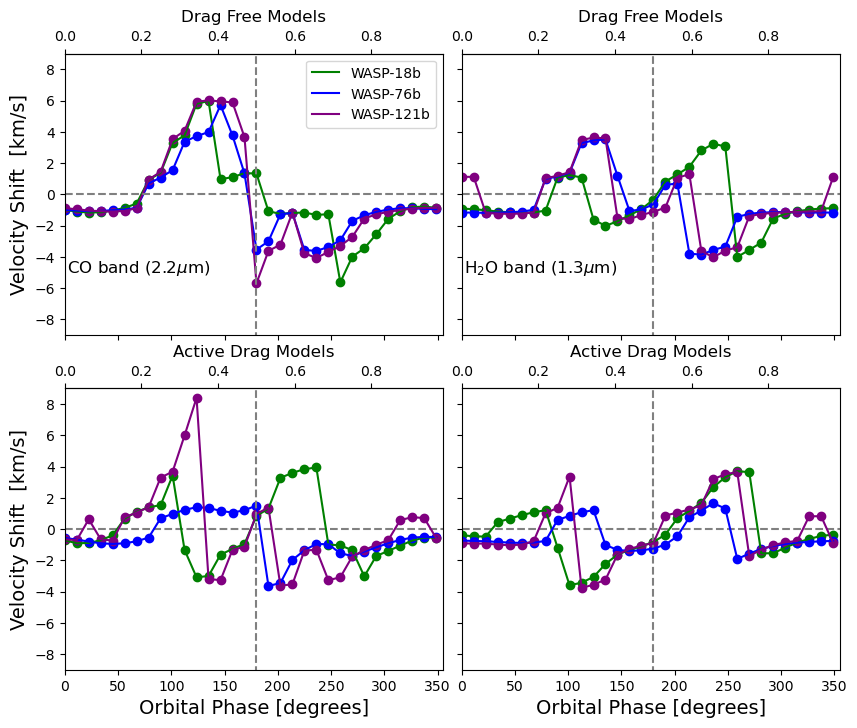}
    \caption{ Net Doppler shifts for both the drag-free and active drag cases for 2.3-2.35$\mu$m and the  1.135-1.355 $\mu$m range (right). The sign of a planet's net Doppler shift before or after secondary eclipse may be indicative of magnetic circulation.   } 
    \label{fig: emissionbothwl}
\end{figure*}

Now, we can begin to compare our models of 3 different UHJs. In Figure \ref{fig: emissionbothwl}, we show the net Doppler shift for our three planets with and without active drag, calculated from high resolution spectra at two different wavelength ranges. 
%The last panel shows the difference between the active and drag free model. For the CO band,  we can see that a models diverge from each other most strongly in the region of time right after first quadrature and just before secondary eclipse. For the water band, we see the largest differences right after secondary eclipse. 
%Although the net difference is planet-dependent, all models show the active drag case being less redshifted/more blueshifted than its corresponding drag-free model during this portion of the orbit. 

Before diving into wavelength dependent behavior, we will examine large scale trends. In each panel of Figure \ref{fig: emissionbothwl} we see a level of scatter between planets, for both the drag free and active drag models. Additionally, this scatter increases for dayside phases. Doppler shifts from the nightside of the planets are on order of a couple km/s regardless of the planet or assumption of drag, resulting in similar behavior when the nightside is primarily in view. The differences between active drag and drag free models here are small, due to the low temperatures of this region resulting in longer timescales that have less of an effect on the atmospheric structure. 

When looking between 1st and 3rd quadrature, the scatter between the models is the largest. There are a variety of processes contributing to this. For WASP-18b, one important consideration is its much higher gravity. The IR photosphere for WASP-76b and WASP-121b lies roughly in the 40-60 mbar range, while for WASP-18, it is closer to 0.1 bar, probing a different part of the planet's circulation. In addition, the equilibrium temperature between these planets vary by nearly 300K, which will result in different temperature and circulation patterns. We additionally note that there appears to be less scatter between models of the same type of drag prescription for the water band. This is likely due to the dissociation of water on the dayside, of which each of these planets are hot enough to do so.  We will explore these differences more thoroughly in the following sections. 
\begin{figure*}
    \centering
    \includegraphics[width=6in]{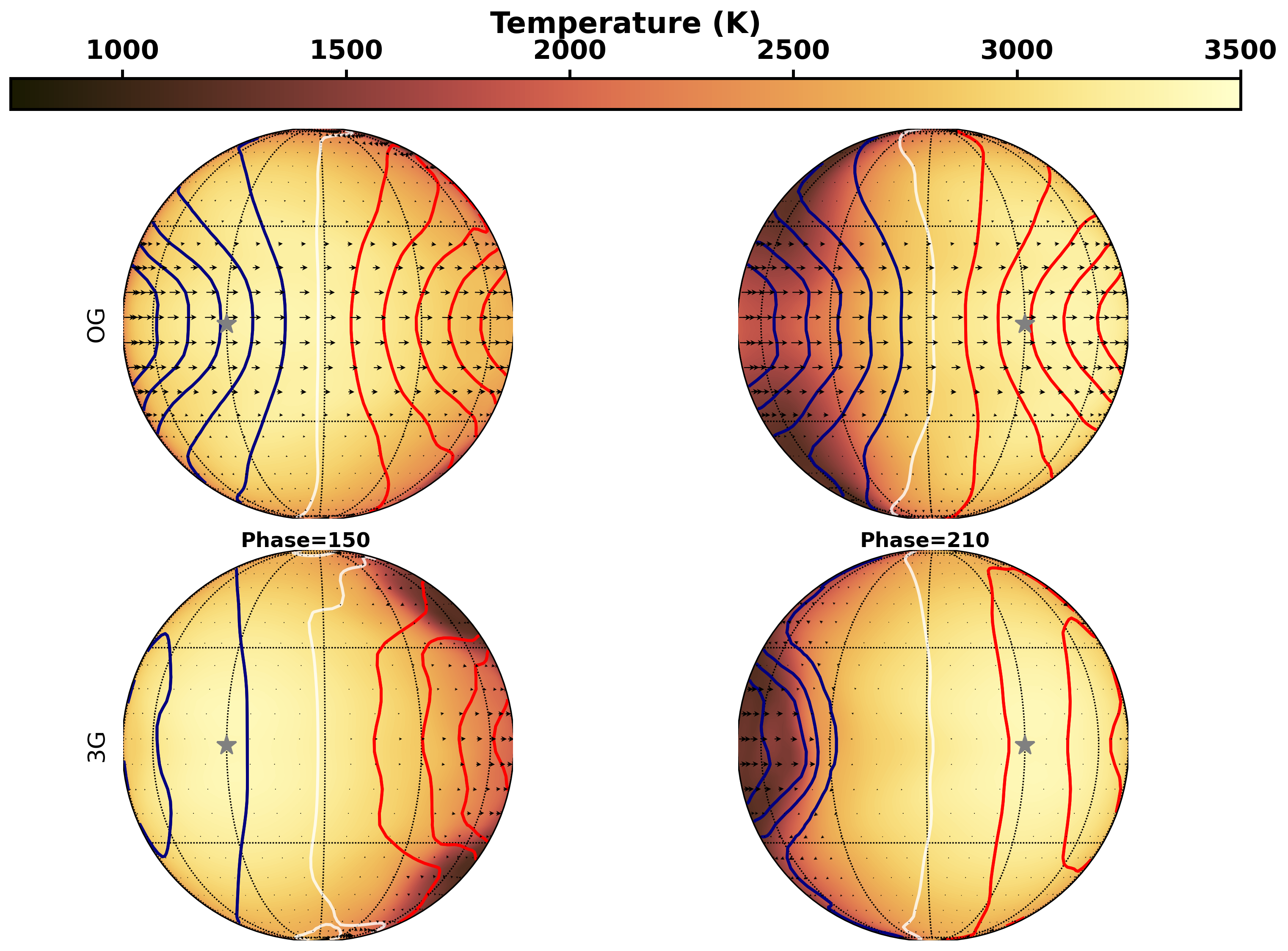}
    \caption{Here we show orthographic projections at a pressure roughly corresponding to the IR photosphere of our models of WASP-121b at phases before and after secondary eclipse. The blue and red lines contours of constant line of sight velocity, in increments of 2 km/s. The grey star shows the location of the substellar point. Prior to secondary eclipse, the 0G model is redshifted while the 3G model is blueshifted. By including active drag, dayside east-west winds are damped, resulting in slower overall wind speeds, but blueshifted winds occupying higher temperature regions at this phase, resulting in a net blueshift prior to secondary eclipse. After secondary eclipse (\textit{right}), both models have a net blueshift, on order of a few km/s. }
    \label{fig:LOS W121}
\end{figure*}

Since HRS is sensitive to line  broadening, we also investigated whether any trends existed in the full width 80\% max of the cross-correlation curves. For each planet, the drag-free models were more broadened than the corresponding active drag for nearly the entire orbit, particularly for dayside phases. This is unsurprising, given that the kinematic MHD approach reduces wind speeds and thus would result in a less broadened cross-correlation curve. 
\subsubsection{CO Behavior}
The scatter between planets means that there is no single universal trend for our active models in emission. However, there are trends near secondary eclipse that can be helpful to find planets that are likely operating under a magnetic circulation regime. We find that, for CO, just prior to secondary eclipse:
\begin{itemize}
    \item A blueshifted spectrum may be indicative of magnetic drag, as 2 out of the 3 active models show this trend. 
    \item A redshifted spectrum is uninformative, as all of the drag free models and one of the active models (WASP-76b) show this trend.
\end{itemize}

After secondary eclipse, all of the models except for WASP-18b are blueshifted. Thus, a redshift after secondary could point to the presence of magnetic drag, but since this was only the case for WASP-18b, it may not be a universal trend for UHJs.  

To understand this behavior, it helps to look at the line of sight velocities at different phases, as shown in Figure \ref{fig:LOS W121}.
%We show the corresponding figures for WASP 76b(Figure \ref{fig:LOS W76}) and WASP18b (Figure \ref{fig:LOS W18}) in the Appendix.
Here, we look at the temperature and wind structure near the IR photosphere before and after secondary eclipse for our models WASP-121b.  Before secondary eclipse, the 3G model is blueshifted while the 0G model is redshifted.  Looking at the 3G model, we notice that the redshifting contours occupy cooler regions than the blueshifted contours, resulting in a net blueshift effect. At first glance, it's not immediately obvious what the net shift of the 0G model should be at this phase. However, as one goes higher in the atmosphere--where the line cores may form---the redshifted winds reach higher speeds and cover more area of the planet, resulting in a net redshift for the drag free model at this phase. After secondary eclipse, both of the models are blueshifted, with the 0G showing a stronger blueshift due to its stronger overall wind speeds. 

We also explored the behavior of a single CO line centered at 2.3132 $\mu$m throughout the planet's orbit. We calculate the equivalent width of this feature of the Doppler off spectra (without broadening from winds and rotation) denoting emission lines as positive and absorption lines as negative. The behaviors for each planet were broadly similar, and we show this equivalent width as a function of phase for WASP-121b in Figure \ref{fig:COW121} and the other two planets in Appendix Figure \ref{fig:COW76W181} where the solid lines denote the drag-free models and the dashed lines show the active drag cases. We additionally show 3 sets of spectra to highlight the range of line shapes present.  Throughout the orbit, the line switches from an absorption feature  to an emission feature  once the dayside is in view and then back to absorption, for both the active and drag free cases. For this CO feature, the drag-free model reaches a stronger peak emission strength and shows features in emission for a larger portion of orbit  compared to the active drag model. Additionally, the drag-free model shows stronger absorption compared to the active drag model for most phases where both models are exhibiting absorption. Similar to \cite{vansluj2023}, we find that the 3D temperature profile has a direct impact on the corresponding line strength.    

\begin{figure}
    \centering
    \includegraphics[width=3.25in]{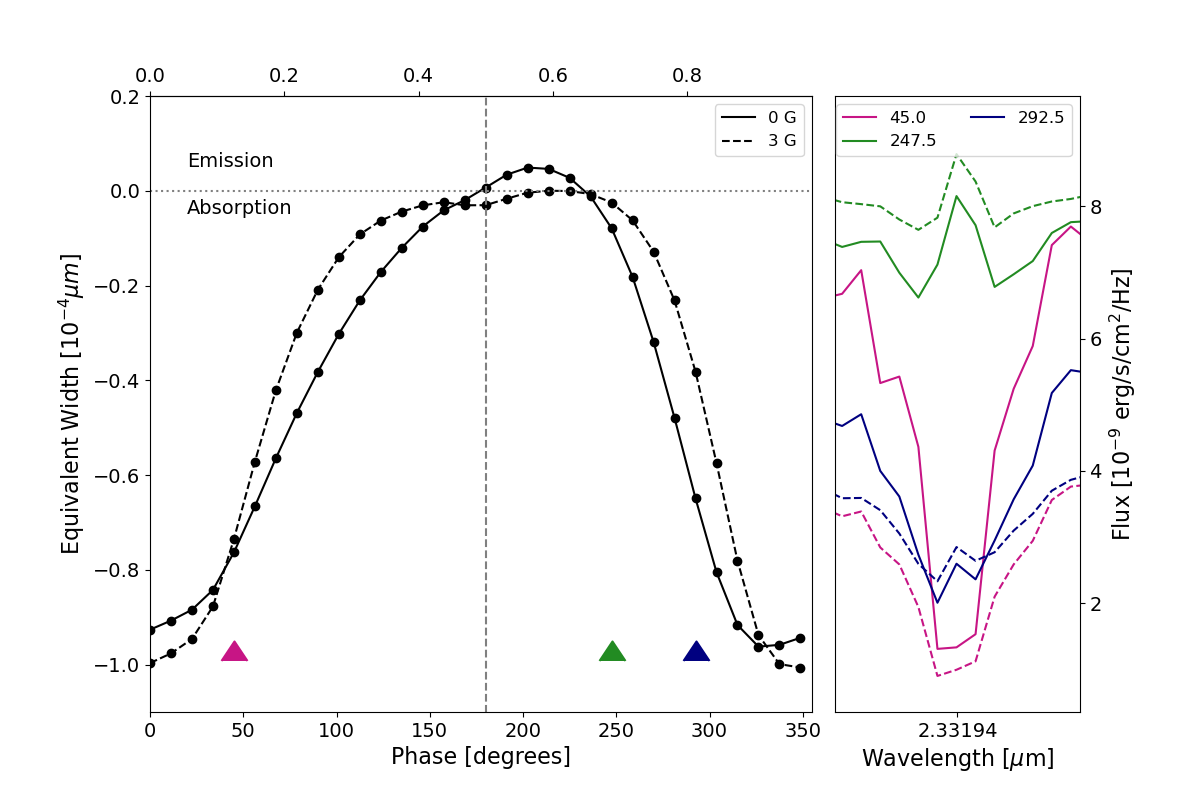}
    \caption{Equivalent width for a single CO line centered at 2.3132 $\mu$m for our drag-free (solid lines) and active drag (dashed lines) models of WASP-121b. The three sets of spectra shown on the right display the range of line profiles present throughout orbit.   Both models show the line in emission while the dayside of the planet is in view. In the case of the spectra at phase 292.5$^{\circ}$, we see that the feature is mostly in absorption, but the line center shows a small amount of emission.   The use of a 3D temperature profile results in varying line strengths throughout a single orbit.  }
    \label{fig:COW121}
\end{figure}

%the drag-free model's wind and temperature structure differs from the active drag case. For the phases shown, the drag-free model consistently has blueshifted contours that reach 10 km/s. The active drag models on the other hand have much weaker blueshifting during this phase range, with maximum contours reaching 6 km/s. The redshifting contours reach up to 8-10 km/s in contrast. However, these redshifting countours are on the cooler part of the planet, which is colder than the drag-free counterpart. For the case of WASP-121b, this results in a net blueshift for our active drag model at the second phase (144) shown in Figure \ref{fig:LOS W121}.

%During this phase range, the dayside of the planet is coming more into view. The dayside winds of our active models are damped strongly in the east-west direction, meaning that the overall wind speed of the active models are consistently weaker. For the active drag case of WASP~121 shown in Figure \ref{fig:LOS W121}, the net Doppler shift goes from redshifted at phase 100, blueshifted at phase 144, and essentially 0 by a phase of 180. During these same phases, the drag-free case is redshifted, redshifted, and then strongly blueshifted. At a phase of 144, the net Doppler shifts between models differ the most, and have different net directions, with the active model being several km/s more blueshifted---an observable trend via phase-resolved spectroscopy with HRS. 

\subsubsection{Water Behavior}
\begin{figure}
    \centering
    \includegraphics[width=3.25in]{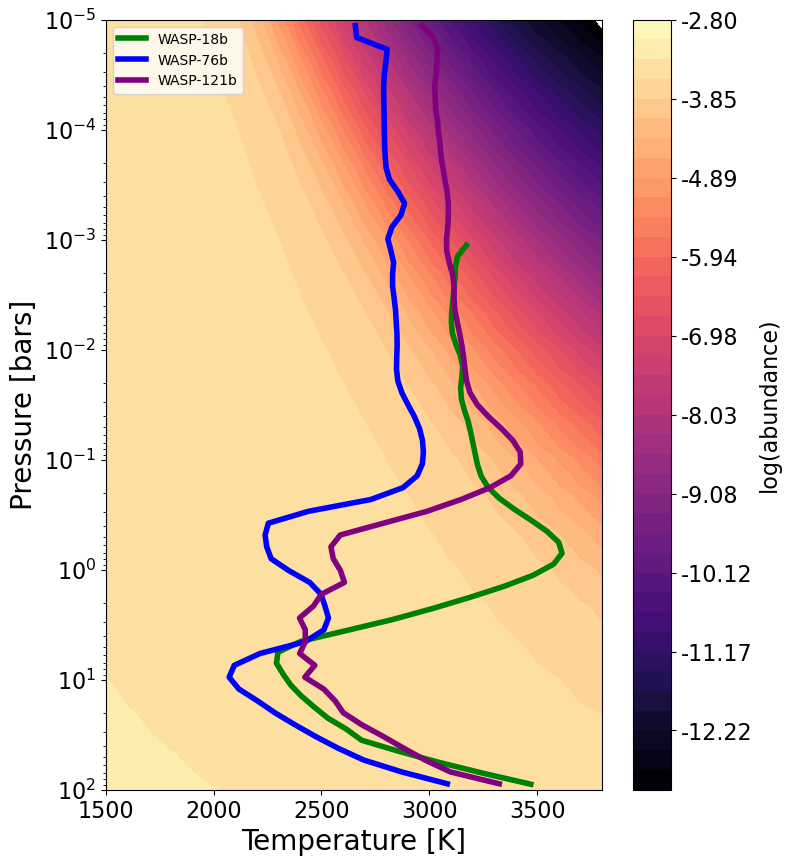}
    \caption{ Here we show H$_{2}$O abundance as a function of temperature and pressure. Overplotted are the substellar profiles from our active drag models. Between the models, we can see that the abundance of water can change by orders of magnitude.  } 
   \label{fig: H2O opacity}
\end{figure}

We also explored a second wavelength range, where water is the dominant opacity and show the corresponding velocity shifts on the right side of Figure \ref{fig: emissionbothwl}. Notably, unlike CO, water is not uniformly abundant around the planet, as shown in Figure \ref{fig: H2O opacity}. In this Figure, we show the abundance of water as a function of temperature and pressure as well as three substellar profiles from our active drag models. We can see that, near the photosphere for WASP-76b and WASP-121 (roughly 40-60mbars), that the two profiles differ in water abundance by a few orders of magnitude. For WASP-18 and its deeper photosphere near 0.1bar, the water abundance is between the two values probed by the photospheres of the two other planets. This is in contrast to CO, which is roughly constant in abundance for the pressures and temperatures probed by the models. 

The varying water abundances add a level of complexity to understanding the net Doppler shifts. Simply looking at temperature and wind maps are no longer sufficient to explain the entirety of the Doppler shift behavior. However, we can still note the trends seen in the net Doppler shifts around secondary eclipse:
\begin{itemize}
    \item Before secondary eclipse, every model at this wavelength is slightly blueshifted. 
    \item Shortly after secondary eclipse, by phase $\sim 225$, all of the active models are redshifted, which may be indicative of active drag. 
    \item After secondary eclipse, the drag free models are more likely to be blueshifted.  The exception to this is the drag-free model of WASP-18b.
\end{itemize}

These trends, combined with the those discussed for CO, help paint the picture of how the magnetic circulation regime can alter phase-resolved high resolution emission observation. 

%For this wavelength, we again see some deviations between our active and drag free models between first quadrature and secondary eclipse, however the largest deviation comes near third quadrature. For this portion of the orbit, our active models are more redshifted than the corresponding drag-free models. This peak difference in velocity shift is planet-dependent, highlighting the influence of water dissociation at various temperatures. The abundance of water varies across the planet, resulting in water lines probing different atmospheric heights. These different pressure levels have different corresponding velocities, which explains why this third quadrature feature was more pronounced in this wavelength range compared to the CO-dominated range.

%We can again look at orthographic projections to understand this effect shown in Figure \ref{fig: W1213rdQuad}. At a phase of 200, both models have roughly the same net Doppler shift very close to zero.  At the middle phase of 225, the more strongly blueshifted regions are restricted to the cool, dim regions near the limb, where they are less influential, making the total shift of the model to be positive. The drag-free case does not have this issue and is blueshifted at this phase. At the final phase shown, the active drag model is more redshifted in comparison.

%\begin{figure*}
    %\centering
    %\includegraphics[width=5in]{W121_comp_orthos_5e-3barsdoubleLOS_thirdquad.png}
    %\caption{  Here we show orthographic projections at a pressure of roughly 5e-3 bars of our models of WASP-121b at multiple phases. The blue and red lines contours of constant line of sight velocity, in increments of 2 km/s. The grey star shows the location of the substellar point. Active drag models are more redshifted near third quadrature than drag free models. } 
   % \label{fig: W1213rdQuad}
%\end{figure*}

\subsection{Trends across Planets: Transmission Spectra}
\begin{figure}
    \centering
    \includegraphics[width=3.5in]{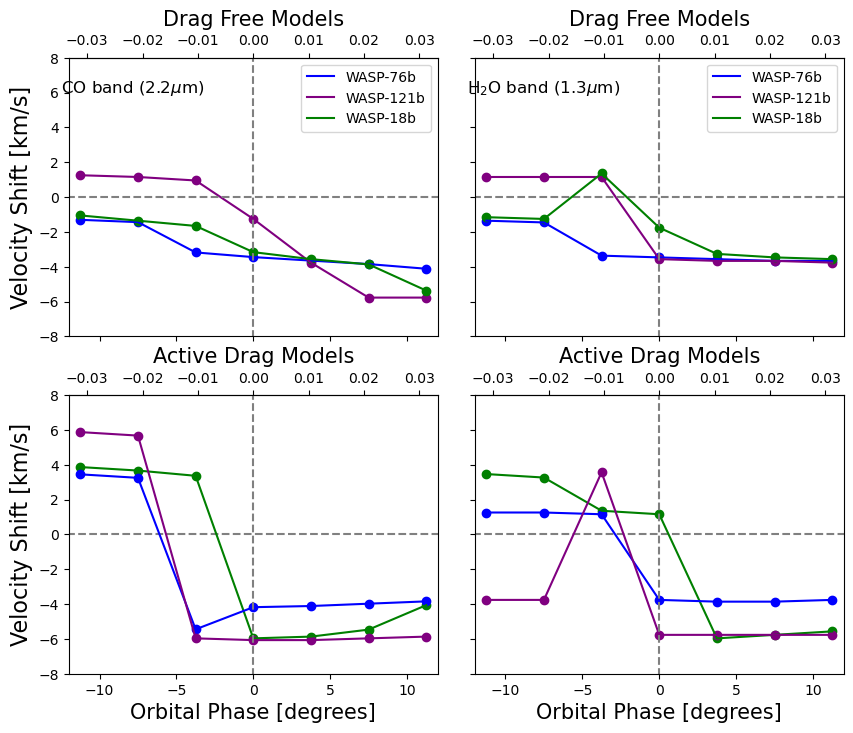}
    \caption{Net Doppler shifts during transmission for the models near 2.2 micron (left) and 1.3 micron (right). For the CO band on the left, we can see the main trend of our active models showing a sharp blueward slope, then flattening out. The drag-free models on the other hand simply become more blueshifted as transit progresses. For the water band, we see different trends between the models in the first half of transit, while  all models flatten out during the second half of transit.    } 
    \label{fig: alltranbothwl}
\end{figure}

We additionally calculated transmission spectra every 3.75 degrees from phases 348.75 to 11.25 degrees  and show the peaks from the cross-correlation analysis in Figure \ref{fig: alltranbothwl} for the CO-dominated and water-dominated wavelength ranges. We identify a consistent trend for these models in the CO band:
\begin{itemize}
    \item The drag free models tend to become more blueshifted as transit progresses with a shallow slope.
    \item Our active drag models show a distinct shape. First, there is a sharp blueshifting slope followed by a flattening of the net Doppler shift. Additionally, the active models are more likely to start the transit redshifted. 
\end{itemize}

 This flattening effect was identified in \cite{Beltz2022b} as it matched the ``bottoming out'' behavior seen in \cite{Ehrenreich2020}. \\
 
 In the water band, we see a variety of trends in the first half of transit , but all models flatten out during the second half. The differences between these wavelengths is again a result of non-uniform water abundances. The uniformity in abundance of CO makes it a better species than water to asses this magnetic circulation feature. Future work should study whether there are additional robust signatures for other species.

%In Figure \ref{fig: alltranbothwl}, we show the velocity shifts for the water-dominated wavelength range on the right. Similar to the emission spectra at this wavelength, the trends are more planet dependent. Again, our active drag models show somewhat steeper slopes during ingress, but this is less pronounced at this wavelength than in CO. The WASP-121b model shows interesting behavior where the velocity shift becomes more redshifted with time for a short window during ingress. We had previously identified this behavior in \citet{Beltz2023}, as a potential signature of the magnetic circulation regime. During egress, the models are not easily distinguishable however. 

%\begin{figure}
 %   \centering
  %  \includegraphics[width=3.5in]{allplanettransmission1_3micron.png}
   % \caption{Doppler shifts during transit for the water dominated wavelength range. Since water abundances differ more strongly throughout the atmosphere for each model, we see more planet-dependant behavior.  } 
    %\label{fig: alltrans1.3micron}
%\end{figure}

\section{Discussion} \label{sec:Discussion}
\subsection{Implications for Observations}
This work represents one of the largest sets of theoretical predictions for UHJs in high resolution spectroscopy. Current spectrographs, such as IGRINS \citep{IGRINS2014}, CRIRES+ \citep{CRIRES+2014}, or PEPSI \citep{Pepsi2015} are well suited to preform these observations. Magnetic fields have still yet to be conclusively found on exoplanets, but UHJs remain perhaps the best planet class to study these effects. Our net Doppler shift transmission predictions---specifically the unique trend seen in our active drag models--- are present among all planets modeled and show similarities to existing phase resolved Doppler shifts \citep{Ehrenreich2020,Borsa2021}. Increasing the number of these phase-resolved transit observations would test this prediction and shed more light into the atmospheric dynamics of these planets. In addition to UHJs, these phase-resolved observations should be preformed on cooler hot Jupiters that are expected to have weaker magnetic effects---potentially indicated by a large hotspot offset---in order to act as a ``drag free'' comparison. 

For the phase-resolved emission spectra, we note that the planet to planet behavior differs more strongly than in transmission. This is likely due to the fact that high resolution emission spectra are more sensitive to temperature differences than transmission spectra and a larger portion of the temperature structure is probed. Of the planets examined here, WASP-121 b abides by all the tentative emission trends discussed, and represents an excellent opportunity to probe the atmospheric dynamics of a planet potentially dominated by magnetic circulation.  

We also note that for both emission and transmission spectra, the trends in the CO band were more distinct than those for the water band. CO remains an excellent tracer species for atmospheric dynamics due to the expected near uniform abundance globally.     

\subsection{Model Limitations and Future Work}
Magnetic effects are notoriously computationally difficult to model in 3D. To derive our active drag timescale, simplifying assumptions were necessary. An important assumption is the kinematic assumption---that is, any magnetic field induced by the atmosphere is small compared to the global field strength. This assumption is valid in regions where the magnetic Reynolds number is less than 1. As discussed in Section \ref{subsec: GCM} and in greater detail in \citep{Beltz2022a}, the hottest regions near the top of models cross this boundary and thus should be treated with caution. These regions levels are much higher than the IR photosphere and are confined to the dayside of the planet so we do not expected them to be influential in our emission spectra too strongly. This may have an impact on our transmission spectra, which probes higher up in the atmosphere, but future models with more sophisticated magnetic prescriptions are needed to gauge the magnitude this effect. Another assumption inherent to our model is the assumption that the planet's magnetic field is a dipole aligned with its axis of rotation. From our own solar system, we see both planets with nearly aligned dipoles (Saturn and Jupiter) and those with magnetic fields that are extremely unaligned or not well described by a dipole (Uranus and Neptune). Currently, astronomers have very little knowledge of exoplanet magnetic field topography, so the assumption of an aligned dipole is a reasonable first step. Future work should explore the effect of tilting the dipole or other more complex multipole magnetic fields. //
These planets were modeled with the assumption of solar metallicity. 
WASP~18b may have a  supersolar metallicity, with \citet{Brogi2023W18} reporting a [M/H] of $1.17 ^{+0.66}_{-1.01}$. For WASP-121b, \citet{Changeat2024} reports a subsolar to solar value: $-0.77 < log(Z) < 0.05$. From \textit{IGRINS} data, \cite{Mansfield2024} retrieves a slightly subolar metallicity of $-0.74 ^{+0.23}_{-.17}$, agreeing with the sub-solar metallicity found in \cite{Gandhi20236UHjs}.  
%\citet{Fu2021} finds a metallicity for WASP-76b that is roughly solar based on \textit{HST} and \textit{Spitzer} data. At the time of this publication, we are unaware of a detailed published metallicity for the WASP-76b.  
%and \cite{Yan2023} reports a C/O ratio that is ``broadly consistent with the value of solar abundance.''
These deviations from solar metallicity would alter the drag strength; a higher metallicity planet will have a higher ion fraction, leading to stronger magnetic drag. Previous GCM studies have found that increasing metallicity results in starlight being deposited higher up in the dayside atmosphere, due to the increased gas opacity \citep{Kataria2015}. This also results in an increased day-night temperature contrast and reduces the hotspot offset of phase curves \citep{Showman2009}.  Seeing that there is a  degeneracy between surface field strength and metallicity,this represents an interesting avenue for future work. 

In HRS, the location of line cores is particularly important, as they set the the shift of our cross-correlation function. These line cores form higher up in the planet's atmosphere than the photosphere and are thus probing different regions than the continuum. These line cores occur at lower pressures in transmission than emission. Our models require an upper boundary that may be deeper than some of these line cores, particularly for the strongest lines. This has the potential to impact our transmission spectra of WASP-18b the most, which has the lowest upper boundary of 10$^{-3}$ bars. Although we are likely not capturing all the line cores in this model, we do note that the observed net Doppler shifts in transmission for this planet shown in Figure \ref{fig: alltranbothwl} are similar to those of the other two planets modeled, potentially indicating this effect is minimal. We do not expect this to be a problem for the emission spectra, as these spectra probe deeper in the atmosphere.  

In order to focus on the effects of our active drag, other physical processes were not included. One of these is the inclusion of clouds and their radiative feedback. Other work with our model has focused on the effects of clouds across temperature and radiative transfer schemes \citep{Harada2021, Roman_2021,Malsky2023}. Given the extreme temperatures of the dayside of these planets, clouds are likely non-existent on the dayside, though may be patchy at the cooler upper latitudes of the nightside \citep{Roman_2021,Komacek2022}. Other work explores the interaction between clouds and our active magnetic drag (Kennedy et al. submitted). 

Another process that plays a role in UHJ atmospheres is that of molecular hydrogen dissociation and recombination. This process works to reduce the day night temperature contrast \citep{Bell_2018} as the molecular hydrogen is dissociated on the dayside. As winds transport the hydrogen around to the nightside, the cooler temperatures allow for the recombination of molecular hydrogen. When included in a GCM, this process  results in a the decrease of the day-night temperature contrast as well as  weaker wind speeds in the equatorial jet \citep{Tan_2019}. Future model development is necessary for exploring the feedback between magnetic drag and molecular hydrogen dissociation.

\section{Conclusion} \label{sec:Conclusion}
This work builds upon our initial exploration of WASP-76b \citep{Beltz2022b,Beltz2023} to identify whether there are signatures of magnetic circulation in high-resolution spectra that are robust across the UHJ population. When our kinematic MHD was applied, all planets showed the emergence of a ``magnetic circulation regime,'' consisting of poleward winds on the dayside and little to no hotspot offset. By combining state of the art numerical models with sophisticated magnetic drag treatment and post-processing techniques, we have begun to characterize the effect of magnetic drag on high resolution spectra. 
Our main findings are as follows:
\begin{itemize}
  
    \item In high resolution emission spectra, we were able to identify Doppler shift trends near secondary eclipse that may be indicative of magnetic effects. For CO, we found that our magnetic models were more likely to be blueshifted prior to secondary eclipse. For water, we found our magnetic models were more likely to redshifted after secondary eclipse. 
    \item  In high resolution transmission spectra we found our active drag shows a different shape in net Doppler shift than the drag free models. The drag free models tend to become more blueshifted as transit progresses. In our active models we see a sharp blueward slope followed by a flattening of the net Doppler shift. This signature is robust for  CO, but isn't found in water features, which have more complex behavior.  
\end{itemize}

This work represents the largest set of kinematic MHD models of ultrahot Jupiters published to date. The observational measurements described here are already feasible with current ground based HRS instruments. (See \citet{Ehrenreich2020,Gandhi2022, Prinoth2023, Wardenier2023} for transmission spectra  and \citet{Pino2022, Herman2022, vansluj2023} for emission spectra.)As more of these instruments come online, it will become increasingly important to consider magnetic effects for objects in this extreme temperature regime. Astronomers have yet to conclusively identify a magnetic field on an exoplanet, and this paper works to bring astronomers closer to that goal and understand atmospheric magnetic effects better for these planets. 

\section{Acknowledgments}
H.B would like to thank Arjun Savel, Elsie Lee, and Lennart van Sluijs for their help preparing this manuscript. The authors also thank the anonymous referee for their work improving the quality of this paper. This work was generously supported by a grant from the Heising-Simons Foundation. Many of the calculations in this paper made use of the Great Lakes High Performance Computing Cluster maintained by the University of Michigan and the Zaratan Computing cluster maintained by the University of Maryland.

\bibliographystyle{aasjournal}
\bibliography{bib.bib}
\section{Appendix}
\begin{figure*}
    \centering
    \includegraphics[width=6in]{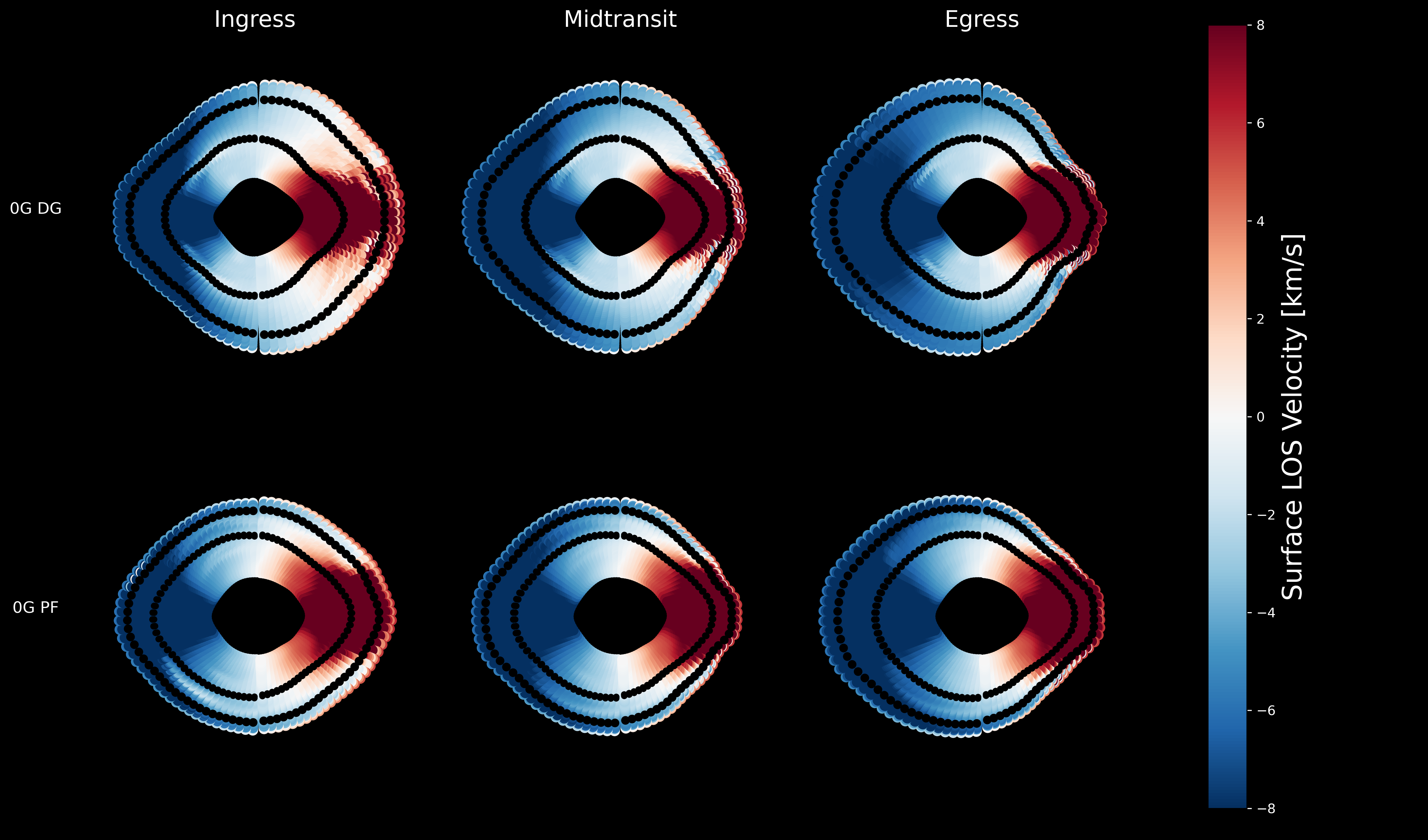}
    \caption{Line of sight velocities for the east and west limbs during ingress, transit, and egress for both the 0G double grey and picket fence models. The similarities between the models result in near identical net Doppler shifts during transit.   }
    \label{fig:LOS W76DG+PFtransit}
\end{figure*}

In Figure \ref{fig:LOS W76DG+PFtransit} we shown the line of sight velocity plots for the drag-free double grey (DG) and picket fence (PF) models of WASP-76b. The presence of the equatorial jet is apparent in both models and is the strongest contributor to the line of sight velocities. The scale height effect arising from the differences in temperature structure between the two models is manifested in the differing spatial extent of the limbs. However, this effect is only minimal here, as the net Doppler shifts during transit for these two models are nearly identical (See Figure \ref{fig: W76DGPF2_micron_trans}).

\begin{figure}
    \centering
    \includegraphics[width=4in]{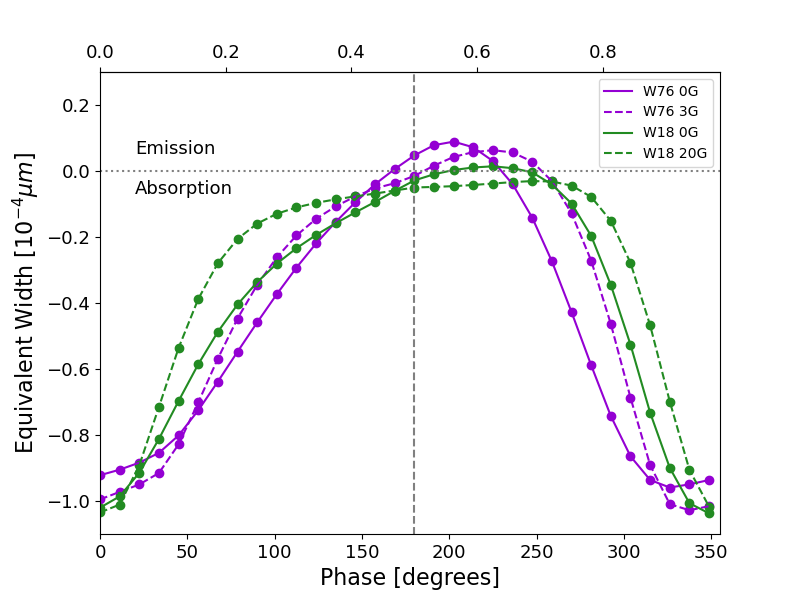}
    \caption{Equivalent width for a single CO feature as a function of phase for our models of WASP-76b and WASP-18b. Broad similarities exist in the behavior of this line, but the differences in the temperature profile result in variations between the active and drag free models as well as between planets.}
    \label{fig:COW76W181}
\end{figure}

Figure \ref{fig:COW76W181} shows the behavior of a CO feature throughout orbit for our WASP-76b and WASP-18b models. The feature shows similar behaviors to those discussed for WASP-121 (see Figure \ref{fig:COW121}) where the drag-free models reach larger equivalent widths in both absorption and emission compared to the active drag models.

%\begin{figure*}
 %   \centering
  %  \includegraphics[width=6in]{W76BPF_comp_orthos_5e-2barsdoubleLOS.png}
   % \caption{W76b orthographic projections.  }
    %\label{fig:LOS W76}
%\end{figure*}

%\begin{figure*}
 %   \centering
  %  \includegraphics[width=6in]{W18_comp_orthos_5e-2barsdoubleLOS.png}
   % \caption{W18b orthographic projections.  }
    %\label{fig:LOS W18}
%%\end{figure*}
\end{document}